\pdfoutput=1

\documentclass[11pt]{article}

\usepackage[pdftex]{graphicx}   
\usepackage{endnotes}            
\usepackage{amsfonts}            
\usepackage{amssymb}             
\usepackage{amsthm}              
\usepackage{amsmath}            
\usepackage{algorithm}           
\usepackage{algorithmic}
\usepackage[letterpaper]{geometry}
\usepackage{rotating}            

\usepackage{setspace}                
\usepackage{xspace}
\newcommand{\BG}{\ensuremath{\mathrm{BG}}\xspace}
\newcommand{\CUG}{\ensuremath{\mathrm{CUG}}\xspace}
\newcommand{\ilogit}{\ensuremath{\mathrm{logit}^{-1}}\xspace}

\newcommand{\binomd}{\ensuremath{\mathrm{Binom}}\xspace}
\newcommand{\Dirichletd}{\ensuremath{\mathrm{Dirichlet}}\xspace}
\newcommand{\CG}{\ensuremath{\mathrm{CG}}\xspace}

\usepackage{natbib}
\bibpunct{(}{)}{;}{a}{,}{;}

\begin{document}

\title{Baseline Mixture Models for Social Networks\thanks{This work was supported under ONR award N00014-08-1-1015, NSF award OIA-1028394, and ARO award W911NF-14-1-0552.  Versions of this work were presented at the 2013 ASA annual meeting, the 2015 Sunbelt conference, the 2016 JSM meeting.}
}

\author{
Carter T. Butts\thanks{Departments of Sociology, Statistics, and EECS, and Institute for Mathematical Behavioral Sciences; University of California, Irvine; Irvine, CA 92697-5100; \texttt{buttsc@uci.edu}}
}
\date{10/10/14}
\maketitle

\begin{abstract}
Continuous mixtures of distributions are widely employed in the statistical literature as models for phenomena with highly divergent outcomes; in particular, many familiar heavy-tailed distributions arise naturally as mixtures of light-tailed distributions (e.g., Gaussians), and play an important role in applications as diverse as modeling of extreme values and robust inference.  In the case of social networks, continuous mixtures of graph distributions can likewise be employed to model social processes with heterogeneous outcomes, or as robust priors for network inference.  Here, we introduce some simple families of network models based on continuous mixtures of baseline distributions.  While analytically and computationally tractable, these models allow more flexible modeling of cross-graph heterogeneity than is possible with conventional baseline (e.g., Bernoulli or $U|man$ distributions).  We illustrate the utility of these baseline mixture models with application to problems of multiple-network ERGMs, network evolution, and efficient network inference.  Our results underscore the potential ubiquity of network processes with nontrivial mixture behavior in natural settings, and raise some potentially disturbing questions regarding the adequacy of current network data collection practices.\\[5pt]
\end{abstract}

\theoremstyle{plain}                        
\newtheorem{axiom}{Axiom}
\newtheorem{lemma}{Lemma}
\newtheorem{theorem}{Theorem}
\newtheorem{corollary}{Corollary}

\theoremstyle{definition}                 
\newtheorem{definition}{Definition}
\newtheorem{hypothesis}{Hypothesis}
\newtheorem{conjecture}{Conjecture}
\newtheorem{example}{Example}

\theoremstyle{remark}                    
\newtheorem{remark}{Remark}

\section{Introduction}

Continuous mixtures of distributions are widely employed in the statistical literature as models for phenomena with highly divergent outcomes; in particular, many familiar heavy-tailed distributions arise naturally as mixtures of light-tailed distributions\footnote{E.g., the $t$ distribution arises as a mixture of Gaussians.} \citep{johnson.et.al:bk:1994,johnson.et.al:bk:1995}, and play an important role in applications as diverse as modeling of extreme values and robust inference \cite{gelman.et.al:bk:2003}.  In the case of social networks, continuous mixtures of graph distributions can likewise be employed to model social processes with heterogeneous outcomes, or as robust priors for network inference.\footnote{Latent variable models such as those of \citet{hoff.et.al:jasa:2002}, \citet{handcock.et.al:jrssA:2007}, or \citet{nowicki.snijders:jasa:2001} can also be thought of as graph mixtures, but are instead designed to model heterogeneity \emph{within} networks.  Our focus here is on models for heterogeneity \emph{across} network realizations.}  Here, we introduce some simple families of network models based on continuous mixtures of baseline distributions.  While analytically and computationally tractable, these models allow more flexible modeling of cross-graph heterogeneity than is possible with conventional baseline (e.g., Bernoulli or $U|man$ distributions \citep{wasserman.faust:bk:1994}).  They are hence attractive as a starting point for the modeling of multiple networks (either cross-sectionally or in series).

The remainder of the paper proceeds as follows.  We introduce mixtures of Bernoulli graphs across expected densities in Section~\ref{sec_denmix}, following with mixtures of $U|man$ graphs across expected dyad frequencies in Section~\ref{sec_manmix}.  In Section~\ref{sec_applications}, we illustrate the utility of these baseline mixture models with application to problems of multiple-network ERGMs, network evolution, efficient network inference.  Finally, we close with some remarks on some implications of our development for data collection practices in the social network community. 

\section{Density Mixtures} \label{sec_denmix}

Let $G$ be a random graph on fixed vertex set $V$, with corresponding adjacency matrix $Y$.  In this section, we make no particular assumptions regarding the directedness of $G$, nor whether loops are permitted, expressing any support constraints by the restriction that $Y$ belong to some set of possible adjacency matrices $\mathcal{Y}$.  Let $e^*$ be the number of edge variables in $G$, i.e., the maximum number of possible edges in $G$ (or, equivalently, the number of distinct elements in $Y$).  Likewise, let $e(Y)$ be the sum of the distinct elements of $Y$, i.e., the number of edges in $G$, with $n(Y)=e^*-e(Y)$ the corresponding number of edges that are unrealized (i.e., null).  If we take all potential edges to occur as independent and identically distributed Bernoulli trials with probability $\delta$, then we are led to the \emph{homogeneous Bernoulli graph} family,
\begin{equation*}
\BG(Y=y|\delta) = \delta^{e(Y)}(1-\delta)^{n(Y)} \mathbb{I}_\mathcal{Y}(y)
\end{equation*}
where $\mathbb{I}_\mathcal{Y}$ is an indicator function for membership in $\mathcal{Y}$.  $\delta$ can be interpreted as the expected density of $Y$, and indeed we can view the Bernoulli graph distribution as a mixture over graphs of fixed density; e.g., if we define the conditional uniform graph distribution
\begin{equation*}
\CUG(Y=y|e) = \binom{e^*}{e}^{-1}\mathbb{I}_\mathcal{Y}(y)
\end{equation*}
in which every permissible graph with $e$ edges is selected with equal probability, then \BG arises as the binomial mixture
\begin{equation*}
\BG(Y=y|\delta) = \sum_{e=1}^{e^*} \CUG(Y=y|e) \binomd(e|e^*,\delta).
\end{equation*}
A well-known consequence of the binomial distribution of $e(Y)$ under the Bernoulli graphs is the sharply peaked behavior of $e(Y)/e^*$ as the graph order $|V|$ becomes large.\footnote{Assuming some basic regularity conditions on $\mathcal{Y}$, e.g., that the number of potential edges in the support is increasing in $|V|$.}  In the limit of $|V|\to \infty$, $e(Y)/e^* \to \delta$ in mean square, with fluctuations that decline as $(e^*)^{-0.5}$.  Since $e^*$ typically scales as $|V|^2$, convergence of the density to its expectation is quite rapid, and large Bernoulli graphs can be viewed as having approximately constant density.  (Indeed, $\BG(\delta)$ and $\CUG(\delta e^*)$ have asymptotically equivalent behavior in increasing order.  See, e.g., \citet{bollobas:bk:2001}.)

The approximately constant density of the Bernoulli graphs is not always a realistic assumption: whether from endogenous social dynamics or population-level variation, the networks observed in a real sample will frequently show considerably greater density variation.  How can we capture this phenomenon while still retaining much of the simplicity of the Bernoulli graph form?  One approach is to represent $Y$ as arising from a hierarchical process in which the expected density, $\delta$, is drawn from some distribution $f$, and $Y$ is then drawn as a Bernoulli graph conditional on $\delta$.  In this case, the pmf of $Y$ becomes a continuous mixture of Bernoulli graphs, i.e.
\begin{align}
\Pr(Y=y) &= \int_0^1 \BG(y|\delta) f(\delta) d\delta \nonumber \\
&= \int_0^1 \delta^{e(y)}(1-\delta)^{e^*-e(y)} f(\delta) d\delta \nonumber \\
&= \int_0^1 \delta^{e(y)}(1-\delta)^{n(y)} f(\delta) d\delta. \label{e_pmf_denmix}
\end{align}

By selecting different forms for $f$, a wide range of distributions can be generated.  Here, we consider the specific case in which $f$ is chosen to be a beta distribution.

\subsection{Beta Density Mixtures} \label{sec_mix_beta}

Following the above discussion, let us take the marginal distribution of density to be beta distributed, i.e. $f(\delta) = \mathrm{Beta}(\delta|\alpha,\beta)$.  Substitution into Equation~\ref{e_pmf_denmix} gives us the corresponding pmf for $Y$:
\begin{align}
\Pr(Y=y|\alpha,\beta) &= \int_0^1 \BG(y|\delta) \mathrm{Beta}(\delta|\alpha,\beta) d\delta \label{e_mixform_beta_den}\\
&= \int_0^1 \delta^{e(Y)}(1-\delta)^{n(Y)} \frac{\Gamma(\alpha+\beta)}{\Gamma(\alpha) \Gamma(\beta)} \delta^{\alpha-1} (1-\delta)^{\beta-1} d\delta \nonumber \\
&= \frac{\Gamma(\alpha+\beta)}{\Gamma(\alpha) \Gamma(\beta)} \int_0^1 \delta^{e(y)+\alpha-1} (1-\delta)^{n(y)+\beta-1} d\delta \nonumber \\
&= \frac{\Gamma(\alpha+\beta)}{\Gamma(\alpha) \Gamma(\beta)} \frac{\Gamma(e(y)+\alpha) \Gamma(n(y)+\beta)}{\Gamma(e^*+\alpha+\beta)}. \label{e_pmf_den}
\end{align}
Gratifyingly, we are left with a closed-form expression involving the ratio of two beta functions ($B(\alpha,\beta)$ and $B(e(y)+\alpha,n(y))$), a value that is readily calculated.  Indeed, the form of Equation~\ref{e_pmf_den} is extremely similar to the pmf of the beta-binomial distribution, to which it is related in the same manner as the relationship between the binomial distribution and the Bernoulli graphs (i.e., $e(Y)$ follows a beta-binomial distribution, with all graphs having the same number of edges being equiprobable).  For this reason, we call this family of Bernoulli graph mixtures the \emph{beta-Bernoulli graphs}.

Although the pmf of Equation~\ref{e_pmf_den} is fairly straightforward to work with, for some purposes it may be more helpfully represented in exponential family (i.e., exponential family random graph, or ERG) form \citep{holland.leinhardt:jasa:1981,frank.strauss:jasa:1986}.  To rewrite the beta-Bernoulli graph in ERG form, we first note that, by definition,
\[
\Pr(Y=y|\alpha,\beta) = \exp\left[ \rho(y,\alpha,\beta) - \log \chi(\alpha,\beta) \right] \mathbb{I}_\mathcal{Y}(y),
\]
where $\rho$ is the ``graph potential'' (a function of the graph, and the model parameters) and $\chi(\alpha,\beta)=\sum_{y'\in \mathcal{Y}}\exp(\rho(y',\alpha,\beta))$ is the normalizing factor that ensures that the resulting quantity is in fact a probability.  Direct inspection of Equation~\ref{e_pmf_den} reveals that we can easily divide the pmf into two factors, one of which depends on $y$ and the other only on the model parameters.  Placing the result in ERG form, we have
\begin{equation}
\Pr(Y=y|\alpha,\beta) = \exp\left[\log\left[\Gamma(e(y)+\alpha) \Gamma(n(y)+\beta)\right] - \log\left[\frac{\Gamma(\alpha) \Gamma(\beta)\Gamma(e^*+\alpha+\beta)}{\Gamma(\alpha+\beta)}\right]\right] \mathbb{I}_\mathcal{Y}(y), \label{e_pmf_den_erg}
\end{equation}
with (curved) potential $\rho(y,\alpha,\beta)=\log\left[\Gamma(e(y)+\alpha) \Gamma(n(y)+\beta)\right]$ and normalizing factor \break$\chi(\alpha,\beta)=\frac{\Gamma(\alpha) \Gamma(\beta)\Gamma(e^*+\alpha+\beta)}{\Gamma(\alpha+\beta)}$.  Although this family is similar to the Bernoulli graphs in having a closed-form normalizing factor, it is \emph{not} an independence model (i.e., edge realizations are not conditionally independent given the rest of the graph).  This is most easily appreciated by considering the conditional probability of an edge being present, which is easily obtained from the exponential family representation:
\begin{align}
\Pr(Y_{ij}=1|Y^c_{ij}=y^c_{ij},\alpha,\beta) &= \ilogit\left[\rho\left(y^+_{ij},\alpha,\beta\right)-\rho\left(y^-_{ij},\alpha,\beta\right)\right]  \nonumber\\
\begin{split}
&=\ilogit\left[ \log\left[\Gamma(e(y^c_{ij})+1+\alpha) \Gamma(n(y^c_{ij})+\beta)\right] \right.\\
& \phantom{=}\quad \left. - \log\left[\Gamma(e(y^c_{ij})+\alpha) \Gamma(n(y^c_{ij})+1+\beta)\right] \right]
\end{split} \nonumber\\
&=\ilogit\left[ \log\left[\frac{\Gamma(e(y^c_{ij})+1+\alpha) \Gamma(n(y^c_{ij})+\beta)}{\Gamma(e(y^c_{ij})+\alpha) \Gamma(n(y^c_{ij})+1+\beta)}\right]  \right] \nonumber\\
&=\left[1+ \exp\left[-\log\left[\frac{e(y^c_{ij})+\alpha }{n(y^c_{ij})+\beta}\right] \right] \right]^{-1} \nonumber\\
&=\frac{e(y^c_{ij})+\alpha}{e(y^c_{ij})+\alpha+n(y^c_{ij})+\beta} \nonumber\\
&=\frac{e(y^c_{ij})+\alpha}{e^*-1+\alpha+\beta}.\label{e_confprob_den}
\end{align}
The full conditionals for this model are thus pleasingly interpretable: the probability of an $i,j$ edge, given the rest of the graph, is equal to the number of other edges in the graph (plus a ``bias'' term, $\alpha$) divided by the number of other edges that are possible (again, plus a ``bias'').  This ratio is effectively an adjusted version of the graph density, and indeed corresponds exactly to the posterior mean estimate of $\delta$ (as defined above) for the Beta model given prior parameters $\alpha,\beta$.  This gives the beta-Bernoulli graphs a ``contagious'' behavioral interpretation (in the language of \citet{coleman:bk:1964}), to which we will return in Section~\ref{sec_contagion}.  Importantly, however, the dependence of the $i,j$ edge probability on $e(y^c_{ij})$ confirms that the beta-Bernoulli graphs are not edgewise independent.  This makes the beta-Bernoulli family a rare example of an edgewise dependent graph distribution that is nevertheless analytically tractable. 

\subsubsection{Mean Degree Parameterization} \label{sec_den_meandeg}

In some settings, it is more natural to consider mixtures over the mean degree than over the density; trivially, the mean degree for an order-$N_V$ graph $y$ is equal to $(N_V-1)\tfrac{e(y)}{e^*}$,\footnote{Since the mean indegree is equal to the mean outdegree in the directed case, we use ``mean degree'' generically to refer to either the undirected mean degree or the mean in/outdegree as required by context.} with expectation $(N_V-1)\delta$.  For the beta-Bernoulli family, this leads to a very simple reparameterization in terms of the expectation and standard deviation of the mean degree.

To see, this, note that $\delta$ under the beta distribution described above has expectation given by
\begin{equation*}
\mathbf{E} \delta = \frac{\alpha}{\alpha+\beta}
\end{equation*}
and variance
\begin{equation*}
\mathrm{Var}(\delta) = \frac{\alpha\beta}{(\alpha+\beta)^2(\alpha+\beta+1)}.
\end{equation*}
Let $\mu_d=(N_V-1) \mathbf{E}\delta$ and $\sigma_d = (N_V-1) (\mathrm{Var}(\delta))^{0.5}$ denote the expectation and standard deviation of the mean degree, respectively.  (These follow immediately from the linear relationship between density and mean degree.)  Solving for parameters $\alpha,\beta$ in terms of these quantities gives us
\begin{gather}
\alpha = \frac{\mu_d}{N_V-1} \left[\frac{\mu_d (N_V-1)}{\sigma_d^2}\left(1-\frac{\mu_d}{N_V-1}\right) -1\right] \label{e_alpha_bydeg}\\
\beta = \left(1-\frac{\mu_d}{N_V-1}\right) \left[\frac{\mu_d (N_V-1)}{\sigma_d^2}\left(1-\frac{\mu_d}{N_V-1}\right) -1\right], \label{e_beta_bydeg}
\end{gather}
a solution which is well-defined whenever $\tfrac{\sigma_d^2}{(N_V-1)^2}< \tfrac{\mu_d}{N_V-1}\left(1-\tfrac{\mu_d}{N_V-1}\right)$ (i.e., $\mathrm{Var}(\delta)<(\mathbf{E}\delta)(1-\mathbf{E}\delta)$, a basic property of the beta distribution).  Substitution of the right-hand sides of equations~\ref{e_alpha_bydeg} and \ref{e_beta_bydeg} for $\alpha$ and $\beta$ (respectively) in equations~\ref{e_mixform_beta_den}--\ref{e_confprob_den} yields what we shall call the \emph{mean degree parameterization} of the beta-Bernoulli graphs.

One obvious application of the mean degree parameterized beta-Bernoulli graphs is to modeling of graph sets with varying $N_V$ but constant expected mean degree.  This is discussed further in section~\ref{sec_ergm_beta}.

\subsection{Simulation and Inference}

One particularly attractive feature of the mixture structure of Equation~\ref{e_pmf_denmix} is the fact it suggests a very direct simulation strategy: simply draw $\delta$ with distribution $f$, and then draw $G\sim \BG(\delta)$.  The former is a one-dimensional continuous random variable with finite range, for which a variety of standard techniques exist \citep[see, e.g.][]{givens.hoeting:bk:2005}.  The latter is a homogeneous Bernoulli graph, which is again easily simulated \citep[e.g.,][]{batagelj.brandes:pre:2005}.  This procedure is exact (so long as its components are exact), and yields independent draws from $G$.  Thus, as a practical matter, density mixtures are much easier to simulate than most other network models with edgewise dependence.

On the inferential side, the closed-form pmf of Equation~\ref{e_pmf_den} would seem to make likelihood-based inference straightforward.  While this is generally true, the beta-Bernoulli family differs from most currently employed network models in that its parameters cannot be identified from a single observation.  In particular, note that for a single graph, $y$ the likelihood has a direction of recession such that $\Pr(Y=y|\alpha,\beta)$ is increasing for $\alpha/(\alpha+\beta)=e(y)/e^*(y), \alpha,\beta \to \infty$.  Taking this limit leads to the Bernoulli graph with expected density equal to the density of $y$, which is intuitively consistent with the notion that, in the absence of variation, one can do no better in a maximum-likelihood sense than to concentrate all mass on the one density value that was observed.  Although a Bayesian approach with appropriate priors on $\alpha$ and $\beta$ will mitigate this problem, the fact remains that effective inference for both parameters depends on multiple graph observations.  As a data model, then, this family is better suited to joint analysis of graph sets (e.g., sampled from a population of graphs) than to single case studies.

\subsubsection{ERGM Inference with Multiple Observations} \label{sec_ergm_beta}

Where multiple realizations from a beta-Bernoulli model are observed, the above-noted problems of identifiability do not apply.  Here, we provide some more detailed results related to inference for the multiple-realization beta-Bernoulli family in ERGM form.  The principal motivation for this approach (as opposed to working with the model in mixture form) is the ease with which the baseline mixture family can be extended by the addition of additional statistics to the graph potential.  While we focus here on the terms related directly to the beta-Bernoulli family, it should thus be emphasized that in practice these terms can be added to others \citep[see, e.g.][for examples]{wasserman.pattison:p:1996,snijders.et.al:sm:2006} to form more complex models.  An empirical example is shown in section~\ref{sec_ergapp}.

Let $\mathbf{Y}=(Y^{(1)},\ldots,Y^{(k)})$ be a collection of $k$ independent random graphs with respective orders $N_V^{(1)},\ldots,N_V^{(k)}$ and edge count maxima $e^{*(1)},\ldots,e^{*(k)}$, for which we observe draws $\mathbf{y}=(y^{(1)},\ldots,y^{(k)})$.  Following equation~\ref{e_pmf_den_erg} may write the joint likelihood of $\mathbf{y}$ in pooled ERGM form as
\begin{align}
\Pr(\mathbf{Y}=\mathbf{y}|\alpha,\beta) &= \prod_{i=1}^k \frac{\exp\left[\log\left[\Gamma\left(e\left(y^{(i)}\right)+\alpha\right) \Gamma\left(n\left(y^{(i)}\right)+\beta\right)\right]\right]}{\frac{\Gamma(\alpha) \Gamma(\beta)\Gamma(e^{*(i)}+\alpha+\beta)}{\Gamma(\alpha+\beta)}} \mathbb{I}_\mathcal{Y}(y^{(i)}) \label{e_ergm_den_pooled}\\
&\propto \exp\left[\sum_{i=1}^k \log\Gamma\left(e\left(y^{(i)}\right)+\alpha\right) + \sum_{i=1}^k \log\Gamma\left(n\left(y^{(i)}\right)+\beta\right)\right]. \label{e_ergm_den_pooled}
\end{align}
This is a curved exponential family on the support of $\mathbf{Y}$, whose canonical form can be obtained via the Taylor series expansions
\begin{gather*}
\log\Gamma\left(e\left(y^{(i)}\right)+\alpha\right) = \log\Gamma(\alpha) + \sum_{j=0}^\infty \frac{\psi^{(j)}(\alpha)}{(j+1)!} e\left(y^{(i)}\right)^{j+1}\\
\log\Gamma\left(n\left(y^{(i)}\right)+\beta\right) = \log\Gamma(\beta) + \sum_{j=0}^\infty \frac{\psi^{(j)}(\beta)}{(j+1)!} n\left(y^{(i)}\right)^{j+1},
\end{gather*}
where $\psi^{(j)}$ is the $j$th derivative of the digamma function.  In an ERGM setting, we may without loss of generality drop constant terms (since these appear in both the numerator and the normalizing factor), leaving us with the canonical pooled ERG form
\begin{align}
\Pr(\mathbf{Y}=\mathbf{y}|\alpha,\beta) &\propto \exp\left[ \sum_{i=1}^k \sum_{j=0}^\infty \frac{\psi^{(j)}(\alpha)}{(j+1)!} e\left(y^{(i)}\right)^{j+1} + \sum_{i=1}^k \sum_{j=0}^\infty \frac{\psi^{(j)}(\beta)}{(j+1)!} n\left(y^{(i)}\right)^{j+1}\right] \nonumber\\
&= \exp\left[ \sum_{j=0}^\infty \frac{\psi^{(j)}(\alpha)}{(j+1)!} \left(\sum_{i=1}^k e\left(y^{(i)}\right)^{j+1}\right) + \sum_{j=0}^\infty \frac{\psi^{(j)}(\beta)}{(j+1)!} \left(\sum_{i=1}^k n\left(y^{(i)}\right)^{j+1}\right)\right]. \label{e_ergm_den_pooled_canonical}
\end{align}
The canonical parameters here are respectively $\psi^{(0)}(\alpha),\ldots$ and $\psi^{(0)}(\beta),\ldots$, with the corresponding canonical statistics being sums of the successive powers of edge and null counts.  As these sums are affinely independent for $k>1$, the model is in general identified.  Although the canonical representation of equation~\ref{e_ergm_den_pooled_canonical} has an infinite number of canonical parameters, these correspond to only two curved parameters ($\alpha$ and $\beta$), and this poses no particular problem for inference.  Unfortunately, however, this series converges too slowly to be of computational use, and hence the canonical form is of theoretical rather than practical interest.\footnote{Expanding about $e^*/2$ yields a series that is much better behaved, but still too slow for use in typical MCMC implementations.}

While modeling this family via its canonical form is impractical, an excellent approximation exists that can be expressed without curved terms.  To obtain it, we begin by observing that $\lim_{x\to \infty} \tfrac{\Gamma(x+z)}{\Gamma(x)x^z}=1$, and that---for reasonably sized graphs---$e(y)$ and $n(y)$ will usually be much larger than $\alpha$ and $\beta$.  Employing the denominator of this expression as an approximation for our curved terms gives us
\begin{gather*}
\log\Gamma\left(e\left(y^{(i)}\right)+\alpha\right) \approx \log\Gamma\left(e\left(y^{(i)}\right)\right) + \alpha \log e\left(y^{(i)}\right), \quad\text{and}\\ 
\log\Gamma\left(n\left(y^{(i)}\right)+\beta\right) \approx \log\Gamma\left(n\left(y^{(i)}\right)\right) + \beta \log n\left(y^{(i)}\right).
\end{gather*}
The resulting model,
\begin{equation}
\Pr(\mathbf{Y}=\mathbf{y}|\alpha,\beta) \propto \exp\left[\sum_{i=1}^k \log\Gamma\left(e\left(y^{(i)}\right)\right) + \sum_{i=1}^k \log\Gamma\left(n\left(y^{(i)}\right)\right) + \alpha \sum_{i=1}^k \log e\left(y^{(i)}\right) + \beta \sum_{i=1}^k \log n\left(y^{(i)}\right) \right], \label{e_ergm_den_approx}
\end{equation}
has two offset terms and two terms that are regular in $\alpha$ and $\beta$ (a considerable simplification).  Since both the offset terms and statistics diverge at zero, it is important to ensure that their sum leads to a normalizable distribution; while divergence towards $-\infty$ is not problematic in this regard (this corresponds to a configuration with zero probability), divergence in a positive direction leads to divergence of the normalizing factor (and an improper distribution).  For the model of equation~\ref{e_ergm_den_approx}, this condition is met so long as $\alpha,\beta\ge 1$.  With this parameter constraint, the model is well-defined, and its distribution approaches that of the beta-Binomial graphs where $e$ and $n$ are large relative to $\alpha$ and $\beta$.

The above model families are appropriate starting points for networks with constant expected density.  In many settings, it is more plausible that the mean degree will be approximately constant; where $N_V$ varies, this requires changes in the expected density.  Fortunately, we can easily reparameterize the above in mean degree form, using the results of section~\ref{sec_den_meandeg}.  Specifically, substituting the expressions of equations \ref{e_alpha_bydeg} and \ref{e_beta_bydeg} for $\alpha$ and $\beta$ in the joint likelihood of equation~\ref{e_ergm_den_pooled} gives us a curved ERGM form for the beta-Bernoulli graphs in terms of mean degree: 
\begin{align}
\begin{split}
\Pr(\mathbf{Y}=\mathbf{y}|\mu,\sigma_d) &\propto \exp\left[\sum_{i=1}^k \log\Gamma\left(e\left(y^{(i)}\right)+\frac{\mu}{N_V^{(i)}-1} \left[\frac{\mu (N_V^{(i)}-1)}{\sigma_d^2}\left(1-\frac{\mu}{N_V^{(i)}-1}\right) -1\right]\right) \right.\\
&\left.+ \sum_{i=1}^k  \log\Gamma\left(n\left(y^{(i)}\right)+\left(1-\frac{\mu}{N_V^{(i)}-1}\right) \left[\frac{\mu (N_V^{(i)}-1)}{\sigma_d^2}\left(1-\frac{\mu}{N_V^{(i)}-1}\right) -1\right]\right)\right]. \label{e_ergm_mdeg_pooled}
\end{split}
\end{align}
Derivation of the canonical and log-approximation forms proceed analogously.  

By adding additional statistics to the potential of equation~\ref{e_ergm_mdeg_pooled}, one can construct model families whose baseline behavior scales realistically with the number of nodes, and which are also overdispersed (relative to a Bernoulli graph) in the same manner as the beta-Bernoulli graphs.  Such properties are of obvious importance when examining populations of networks from organizational or other settings in which the number of actors can vary substantially.  We consider just such an example in Section~\ref{sec_ergapp}.

\section{Dyad Frequency Mixtures} \label{sec_manmix}

While density mixtures are natural for undirected graphs, baseline models for directed graphs often consider variation in the tendency for edges to be reciprocated (known e.g., as asymmetry, local hierarchy, or reciprocity).  The density and reciprocity are jointly captured by the \emph{dyad census} \citep{wasserman.faust:bk:1994}, the number of dyads (unordered vertex pairs) whose induced subgraphs are mutual (both ties present), asymmetric (one tie present), or null (no ties present).  For graph adjacency matrix $Y$, let us designate $M(Y)$ to be the count of mutuals, $A(Y)$ the count of asymmetrics, and $N(Y)$ the count of nulls, with $D^*(Y)=M(Y)+A(Y)+N(Y)$ the total number of dyads.  Where $Y$ is undirected, $A(Y)$ is by definition equal to 0; in the general case, however, all three types may be present in varying frequencies.  As dyads are non-intersecting, it is natural to consider a baseline model in which each dyad state is chosen independently, in a categorical\footnote{We refer to the multivariate analog of the Bernoulli distribution as the \emph{categorical} distribution, reserving \emph{multinomial} for the count distribution of a set of iid categorical draws.} analog to the homogeneous Bernoulli graph.  This homogeneous ``categorical graph'' can be parameterized as follows: 
\begin{equation*}
\CG(Y=y|m,a,n) = \left(\frac{1}{2}\right)^{A(y)} m^{M(y)} a^{A(y)} n^{N(y)} \mathbb{I}_\mathcal{Y}(y),
\end{equation*}
with the dyad census statistics as defined above, and the parameters $m$, $a$, and $n$ corresponding to the probability of obtaining a mutual, asymmetric, or null dyad, respectively.  (Note that these parameters have only two degrees of freedom, as e.g. $n=1-a-m$, but are shown here in explicit form for substantive clarity.)  This family is generally known as the ``$U|man$'' distribution (shorthand for ``uniform given mutual, asymmetric, and null dyad frequencies''), and is a widely employed baseline model for directed graphs \citep[see, e.g.][]{wasserman.faust:bk:1994}.  Although it is convenient to specify this model directly in terms of dyad frequencies (as done here), it should be noted that we can easily recover other familiar local properties by appropriate transformations of the dyad census.  For instance, the expected density of $\CG(Y=y|m,a,n)$ is $m+a/2$ (since dyads are iid, and each carries at most two edges), the expected edgewise reciprocity is $2m/(2m+a)$, etc.

As with the Bernoulli graph, a conditional uniform variant of the $U|man$ family also exists; and, as with the Bernoulli graph, the sharply peaked behavior of the multinomial leads to asymptotic dyad frequencies strongly concentrated around their expectations.  To avoid this behavior, it is natural to consider a mixture of categorical graphs, this time expressed in terms of dyad frequencies:
\begin{align}
\Pr(Y=y) &= \int_0^1 \int_0^{1-m} \CG(Y=y|m,a,1-m-a) f(m,a,1-m-a) dm da \nonumber\\
&= \int_0^1 \int_0^{1-m} \left(\frac{1}{2}\right)^{A(y)} m^{M(y)} a^{A(y)} n^{N(y)} f(m,a,1-m-a) dm da. \label{e_dyadmix}
\end{align}
$f$ in this case is taken to be the underlying dyad frequency distribution, with $Y$ drawn conditional on the frequency parameters.  Selecting different forms for $f$ leads to graph distributions with different tendencies towards high or low density and/or reciprocity, and different levels of joint variation in these characteristics.  Here, we will examine a particular choice of $f$: the Dirichlet family, which serves as the direct analog of the beta distribution in the Bernoulli case.  Our discussion likewise proceeds in parallel to our development in Section~\ref{sec_mix_beta}.

\subsection{Dirichlet Dyad Frequency Mixtures}

In the beta-Bernoulli case, our mixture was in terms of a single parameter (or, equivalently, two parameters with one degree of freedom).  In this case, we have three parameters in the $(0,1)$ interval that must sum to 1 (i.e., a distribution on the standard 2-simplex).  A simple and familiar distribution with this support is the Dirichlet, the generalization of the beta distribution in $k$ dimensions.  Here, we will write $\Dirichletd(m,a,n|\alpha,\beta,\gamma)$ to refer to the Dirichlet pdf evaluated at $(m,a,n)$ with parameter vector $(\alpha,\beta,\gamma)$, bearing in mind our constraint that $m+a+n=1$.  Intuitively, $\alpha$, $\beta$, and $\gamma$ respectively reflect the tendency of the distribution to place greater weight on $m$, $a$, and $n$; indeed, the expectation of $m$ is $\alpha/(\alpha+\beta+\gamma)$, the expectation of $a$ is $\beta/(\alpha+\beta+\gamma)$, etc.  The sum $\alpha+\beta+\gamma$ controls the concentration of the distribution about its mean, and indeed has a well-known Bayesian interpretation in terms of a ``prior sample size'' \citep{gelman.et.al:bk:2003} in inferential contexts.  For the present, let us take these parameters to be given.  We then define the \emph{Dirichlet-categorical graph} via the mixture
\begin{align}
\Pr(Y=y|\alpha,\beta,\gamma) &= \int_0^1\int_0^{1-m}\CG(y|m,a,n) \Dirichletd(m,a,n|\alpha,\beta,\gamma) dm da. \nonumber \\
\intertext{Substitution from the definitions of these respective distributions leads to}
\begin{split}
&= \int_0^1\int_0^{1-m} \left[ \left(\frac{1}{2}\right)^{A(y)} m^{M(y)} a^{A(y)} (1-m-a)^{N(y)} \right.\\
&\left.\phantom{=\int_0^1\int_0^{1-m}\quad}\times \frac{\Gamma\left(\alpha+\beta+\gamma\right)}{\Gamma(\alpha)\Gamma(\beta)\Gamma(\gamma)} m^{\alpha-1} a^{\beta-1} (1-m-a)^{\gamma-1}\right] dm da
\end{split}  \nonumber\\
\intertext{which factors as}
\begin{split}
&= \left(\frac{1}{2}\right)^{A(y)} \frac{\Gamma\left(\alpha+\beta+\gamma\right)}{\Gamma(\alpha)\Gamma(\beta)\Gamma(\gamma)}\\ 
&\phantom{=\quad} \times \int_0^1\int_0^{1-m} m^{M(y)+\alpha-1} a^{A(y)+\beta-1} (1-m-a)^{N(y)+\gamma-1} dm da.
\end{split} \nonumber\\ 
\intertext{Finally, we note that the integral in question corresponds to the multinomial beta function, with parameters $M(y)+\alpha$, $A(y)+\beta$, and $N(y)+\gamma$.  Substituting the latter (in its gamma function expression) leaves us with}
&=\left(\frac{1}{2}\right)^{A(y)} \frac{\Gamma\left(\alpha+\beta+\gamma\right)}{\Gamma(\alpha)\Gamma(\beta)\Gamma(\gamma)} \frac{\Gamma(M(y)+\alpha)\Gamma(A(y)+\beta)\Gamma(N(y)+\gamma)}{\Gamma\left(D^*(y)+\alpha+\beta+\gamma\right)}. \label{e_dmix_pmf}
\end{align}
Like the beta-Bernoulli graph, the pmf of the Dirichlet-categorical graph has a closed-form expression involving a ratio of beta functions.  In this case, we also have a combinatorial factor arising from the multiplicity of the asymmetric dyads (each of which has two distinct orientations).  Regardless, the pmf of Equation~\ref{e_dmix_pmf} is easily computed, and its form is relatively simple.

As with the beta-Bernoulli graph, it can be helpful to consider the exponential family representation of the Dirichlet-categorical family.  Returning to our basic definition, we have
\begin{align*}
\Pr(Y=y|\alpha,\beta,\gamma) &= \exp\left[ \rho(y,\alpha,\beta,\gamma) - \chi(\alpha,\beta,\gamma) \right] \mathbb{I}_\mathcal{Y}(y), \\
\intertext{and hence by substitution from Equation~\ref{e_dmix_pmf}}
&= \exp\left[\log \frac{\Gamma(M(y)+\alpha)\Gamma(A(y)+\beta)\Gamma(N(y)+\gamma)}{2^{A(y)}\Gamma\left(D^*(y)+\alpha+\beta+\gamma\right)} - \log \frac{\Gamma(\alpha)\Gamma(\beta)\Gamma(\gamma)}{\Gamma\left(\alpha+\beta+\gamma\right)} \right].
\end{align*}
It therefore follows that $\rho(y,\alpha,\beta,\gamma)=\log\tfrac{\Gamma(M(y)+\alpha)\Gamma(A(y)+\beta)\Gamma(N(y)+\gamma)}{2^{A(y)}\Gamma\left(D^*(y)+\alpha+\beta+\gamma\right)}$ and $\chi(\alpha,\beta,\gamma)=\log \tfrac{\Gamma(\alpha)\Gamma(\beta)\Gamma(\gamma)}{\Gamma\left(\alpha+\beta+\gamma\right)}$.  From this we may derive the conditional probability of a given edge.  By the properties of the exponential family, we have
\begin{align}
\Pr(Y_{ij}=1|Y^c_{ij}=y^c_{ij},\alpha,\beta,\gamma) &= \ilogit\left[\rho\left(y^+_{ij},\alpha,\beta,\gamma\right)-\rho\left(y^-_{ij},\alpha,\beta,\gamma\right)\right], \nonumber\\
\intertext{and hence}
&= \left[1 + \exp\left[ \log \tfrac{2^{A(y^+_{ij})}\Gamma(M(y^-_{ij})+\alpha)\Gamma(A(y^-_{ij})+\beta)\Gamma(N(y^-_{ij})+\gamma)}{2^{A(y^-_{ij})}\Gamma(M(y^+_{ij})+\alpha)\Gamma(A(y^+_{ij})+\beta)\Gamma(N(y^+_{ij})+\gamma)} \right]\right]^{-1} \nonumber\\
&=\left[1 + y_{ji} \frac{\frac{1}{2}\left(A(y^+_{ij})+\beta\right)}{M(y^-_{ij})+\alpha} + \left(1-y_{ji}\right) \frac{N(y^+_{ij})+\gamma}{\frac{1}{2}\left(A(y^-_{ij})+\beta\right)}\right]^{-1} \nonumber\\
&=\begin{cases} \frac{M(y^-_{ij})+\alpha}{M(y^-_{ij})+\alpha+\frac{1}{2}\left(A(y^+_{ij})+\beta\right)} & \text{if } y_{ji}=1\\ \frac{\frac{1}{2}\left(A(y^-_{ij})+\beta\right)}{\frac{1}{2}\left(A(y^-_{ij})+\beta\right)+N(y^+_{ij})+\gamma} & \text{if } y_{ji}=0 \end{cases}. \label{e_dmix_condprob}
\end{align}
Note that the expression of Equation~\ref{e_dmix_condprob} depends upon the state of the $(j,i)$ edge -- this is true for any family allowing dependence within dyads (including $U|man$), and is a manifestation of the capacity of the model to show tendencies towards or away from reciprocity.  However, a closer inspection of this expression shows that it depends also on edges outside the dyad.  In particular, the terms involved correspond directly to the counts of mutual, asymmetric, and/or null dyads other than $\{i,j\}$.  Roughly speaking, edges in the Dirichlet-categorical model conditionally occur in ways that are consistent with or reinforce existing dyad frequencies: when there are more mutuals in the population, formation of mutuals is favored, more asymmetric relationships favor the creation/retention of asymmetry, etc.  This is another instance of ``contagious'' behavior on par with that observed in the beta-Bernoulli family, where the ``contagion'' here is at the level of dyadic relationships (rather than edges).  We consider this further in section~\ref{sec_contagion}.

\subsubsection{Mean Non-null Degree/Reciprocity Parameterization} \label{sec_dcat_meandeg}

In analogy to the beta-Bernoulli case, the base Dirichlet-categorical graph parameterization leads to mixtures with constant expected density and degree of symmetry (i.e., dyadic reciprocity); since this is often inappropriate for graph sets with varying size, we may therefore ask whether this family can be parameterized in terms of mean degree.  Although the presence of a reciprocity term complicates matters, we can nevertheless develop a very similar parameterization of the Dirichlet-categorical family.

To begin, we define a \emph{non-null} interaction between $i$ and $j$ as the state in which at least one of the $(i,j),(j,i)$ edges exists.  Since this occurs when the $\{i,j\}$ dyad is either mutual or asymmetric, the corresponding probability under the $U|man$ family is equal to $m+a$.  Likewise, we say that $i$ and $j$ have a mutual interaction if $\{i,j\}$ is a mutual dyad, which occurs with probability $m$.  Define the \emph{reciprocation rate}, $r$, as the probability that an arbitrary non-null dyad is mutual (i.e., $r=m/(m+a)$).  In many settings, it may be useful to model the expected number of non-null interactions per vertex (the mean ``non-null degree,'' $\mu_{nnd}$) and the expected number of reciprocal interactions per vertex (the mean ``mutual degree,'' $\mu_{md}$) as constant in $N_V$.  Trivially, these are given by $\mu_{nnd}=(N_V-1)\mathbf{E}(m+a)$ and $\mu_{md}=(N_V-1)\mathbf{E}m=r \mu_{nnd}$.  Specification of $\mu_{nnd}$ and $r$ leaves one parameter to account for dispersion.  Here, we employ the standard deviation of the non-null degree, $\sigma_{nnd}=(N_v-1)(\mathrm{Var} (m+a))^{0.5}$, for this purpose.  This parameterization is then very close to the mean degree parameterization of the beta-Bernoulli family, with the non-null degree taking the place of degree (intuitively, the number of alters with whom ego has \emph{some} interaction) and the added property that some fixed fraction of non-null interactions are mutual (equivalently, that the expected number of mutual ties per node is fixed).

To solve for our Dirichlet parameters ($\alpha,$ $\beta$, and $\gamma$) in terms of the above, we exploit the fact that the distribution of $m+a$ under the Dirichlet distribution is distributed $\mathrm{Beta}(\alpha+\beta,\gamma)$.  (This follows from the standard ``collapseability'' property of the Dirichlet, and the fact that a two-parameter Dirichlet distribution is equivalent to the beta distribution.)  Likewise, the marginal distribution of $m$ is $\mathrm{Beta}(\alpha,\beta+\gamma)$.  Using the beta expectation and variance, and substituting from the above (as per Section~\ref{sec_den_meandeg}), then gives us
\begin{gather}
\alpha = \frac{r \mu_{nnd} \left(\mu_{nnd}\left(N_v-1-\mu_{nnd}\right)-\sigma_{nnd}^2\right)}{(N_V-1)\sigma_{nnd}^2} \label{eq_dcat_mdeg_alpha}\\
\beta = \frac{(1-r) \mu_{nnd} \left(\mu_{nnd}\left(N_v-1-\mu_{nnd}\right)-\sigma_{nnd}^2\right)}{(N_V-1)\sigma_{nnd}^2} \label{eq_dcat_mdeg_beta}\\
\gamma = \frac{\left(N_V-1-\mu_{nnd}\right) \left(\mu_{nnd}\left(N_v-1-\mu_{nnd}\right)-\sigma_{nnd}^2\right)}{(N_V-1)\sigma_{nnd}^2}. \label{eq_dcat_mdeg_gamma}
\end{gather} 
\noindent
The right-hand sides of Equations~\ref{eq_dcat_mdeg_alpha}--\ref{eq_dcat_mdeg_gamma} can be substituted for $\alpha,$ $\beta,$ and $\gamma$ to obtain a Dirichlet-categorical graph family defined in terms of $\mu_{nnd}$, $r$, and $\sigma_{nnd}$.  We refer to this as the \emph{mean non-null degree/reciprocity} parameterization.  This parameterization is well-defined (given that $0<\mu_{nnd}<N_V-1)$ whenever $\mu_{nnd}(N_v-1-\mu_{nnd})>\sigma_{nnd}^2$, 

\subsection{Simulation and Inference}

As for the density mixtures, dyad frequency mixtures of the from in Equation~\ref{e_dyadmix} are particularly easy to simulate: one simply draws the mutual, asymmetric, and null dyad rates from $f(m,a,1-m-a)$, and then draws the graph from the appropriate $U|man$ distribution (a simple problem, with standard implementations).  For the Dirichlet-categorical family, the full conditionals of Equation~\ref{e_dmix_condprob} can also be used for simulation via Gibbs sampling; as this is slow and inexact, the two-stage method is to be preferred in most settings.

Inferential issues pertaining to the Dirichlet-categorical graphs are similar to those affecting the beta-Bernoulli family.  In particular, model parameters cannot be identified with fewer than three observations, since independent variation in both asymmetric and null rates is needed for the MLE to be well-defined.  Given the requisite number of graphs, estimation via MLE or Bayesian methods is straightforward (using the likelihood of Equation~\ref{e_dmix_pmf}).

\subsubsection{ERGM Inference for the Dirichlet-categorical Graphs} \label{sec_ergm_dmix}

Pooled ERGM inference for Dirichlet-categorical graphs with multiple observations parallels the development of section~\ref{sec_ergm_beta}, and we thus treat it only briefly here.  Adapting equation~\ref{e_dmix_pmf} to the multiple observation case, we have 
\begin{multline}
\Pr(\mathbf{Y}=\mathbf{y}|\alpha,\beta,\gamma) \propto \exp\left[ \sum_{i=1}^k \log \Gamma\left(M\left(y^{(i)}\right)+\alpha\right) + \sum_{i=1}^k \log \Gamma\left(A\left(y^{(i)}\right)+\beta\right) \right. \\
\left. + \sum_{i=1}^k \log \Gamma\left(N\left(y^{(i)}\right)+\gamma\right) - \log 2 \sum_{i=1}^k A\left(y^{(i)}\right)  \right]; \label{eq_dcat_curved}
\end{multline}
this is a curved ERGM that clearly resembles that of the beta-Bernoulli family, with the primary difference (other than the number of terms) being the presence of an offset term related to the number of asymmetric dyads.  The corresponding canonical form can be obtained by employing the series expansion of $\log\Gamma(x)$, and is also similar to the beta-Bernoulli case:
\begin{multline}
\Pr(\mathbf{Y}=\mathbf{y}|\alpha,\beta,\gamma) \propto \exp\left[ \sum_{i=1}^k \sum_{j=0}^\infty \frac{\psi^{(j)}(\alpha)}{(j+1)!} M\left(y^{(i)}\right)^{j+1} + \sum_{i=1}^k \sum_{j=0}^\infty \frac{\psi^{(j)}(\beta)}{(j+1)!} A\left(y^{(i)}\right)^{j+1} \right.\\
\left.+ \sum_{i=1}^k \sum_{j=0}^\infty \frac{\psi^{(j)}(\gamma)}{(j+1)!} N\left(y^{(i)}\right)^{j+1} - \log 2 \sum_{i=1}^k A\left(y^{(i)}\right) \right]. \label{eq_dcat_canon}
\end{multline}
The canonical statistics for this family are thus the powers of the mutual, asymmetric, and null dyad counts, together with an offset term for the total number of asymmetric dyads.

As in the beta-Bernoulli case, the power series form of Equation~\ref{eq_dcat_canon} is impractical for inference.  Applying the log gamma approximation used in Section~\ref{sec_ergm_beta} to the curved ERGM form of Equation~\ref{eq_dcat_curved} gives us a more practical approximating model family,
\begin{multline}
\Pr(\mathbf{Y}=\mathbf{y}|\alpha,\beta,\gamma) \propto \exp\left[ \sum_{i=1}^k \log \Gamma\left(M\left(y^{(i)}\right)\right) + \sum_{i=1}^k \log \Gamma\left(A\left(y^{(i)}\right)\right) + \sum_{i=1}^k \log \Gamma\left(N\left(y^{(i)}\right)\right) \right. \\
\left. - \log 2 \sum_{i=1}^k A\left(y^{(i)}\right)+ \alpha \sum_{i=1}^k  \log M\left(y^{(i)}\right) + \beta \sum_{i=1}^k \log A\left(y^{(i)}\right) + \gamma \sum_{i=1}^k \log N\left(y^{(i)}\right)   \right], \label{eq_dcat_apx}
\end{multline}
\noindent
with the first four terms being offsets, and the last three terms linear in the model parameters.  In analogy to the approximate beta-Bernoulli family, this distribution is well-defined so long as $\alpha,\beta,\gamma>1$, with the Dirichlet-categorical limit arising as the dyad statistics become large relative to the model parameters.

A mean non-null degree/reciprocity parameter version of the above can be obtained via substitution from Equations~\ref{eq_dcat_mdeg_alpha}--\ref{eq_dcat_mdeg_gamma}; we omit them here for reasons of brevity.   As with the beta-Bernoulli family, it should be noted that the above ERGM forms can be employed as starting points for the construction of more complex model families, with the latter created by adding additional dependence and/or heterogeneity terms.  Simulation and inference can then be performed using standard techniques \citep[e.g.][]{snijders:joss:2002,hunter.et.al:jss:2008}.  An example of such an analysis is presented in section~\ref{sec_ergapp}.

\section{Applications} \label{sec_applications}


Baseline mixture models have a wide range of uses, both theoretical and methodological.  Here, we consider three: the use of graph mixtures as baselines for ERG models; the use of graph mixtures to study social processes in which social influence affects tie formation; and the use of graph mixtures as weakly informative network priors for Bayesian inference.  All three cases illustrate the power of mixture models to represent graph distributions with widely divergent outcomes, and secondary illuminate important issues involving data collection in support of social network analysis.

\subsection{Graph Mixtures as Multiple-Network ERGM Baselines} \label{sec_ergapp}

In Sections~\ref{sec_ergm_beta} and \ref{sec_ergm_dmix}, we showed that the beta-Binomial and Dirichlet-categorical graphs can be used as baselines for multiple-graph ERGMs; these families provide a simple means of accounting for cross-graph variation in density and/or reciprocity, with models that can be fit using standard techniques.  Here, we illustrate this use of baseline mixture models via an analysis of perceived power relationships in the Urban Communes Data Set \citep{zablocki:bk:1980} (UCDS).  The portion of the UCDS employed here consists of dyadic and individual attribute data from 61 urban communes, whose members were surveyed during 1974.  Specifically, we examine the ``Power'' relation employed by \citet{butts:sm:2011a}, in which an edge exists from individual $i$ to individual $j$ if $i$ or $j$ claims that $i$ exercises power over $j$ in their relationship.  \citet{butts:sm:2011a} employed a Bayesian meta-analytic technique to examine net tendencies towards or away from reciprocity and hierarchy in these networks, finding an overall tendency towards asymmetry (net of density) that varied in strength across communes.  Such variation suggests the use of graph mixtures as a useful baseline for modeling these graphs.

Prior work on the UCDS \citep[e.g.,][]{martin:ajs:2002,martin.fuller:spq:2004,martin:ajs:2005} has shown that perceptions of power exercise are related to a variety of personal and group factors; likewise, studies of dominance in human and other systems \citep[e.g.,][]{chase:asr:1980,chase.et.al:pnas:2002} suggest that power structures within groups should show an endogenous tendency towards transitive (and away from cyclic) closure.  Such mechanisms can be examined simultaneously using ERGMs, but here a straightforward analysis is complicated by the nature of the data: with group sizes ranging from 4 to 26 (mean 9.7), many of the UCDS networks are too small to be modeled independently.  One obvious alternative the use use of pooled ERGMs to jointly estimate parameters across models.  By incorporating mixture terms (sections~\ref{sec_ergm_beta}, \ref{sec_ergm_dmix}) in place of the usual edge term, we can efficiently estimate parameters while still allowing for variation in baseline density and/or reciprocity.

We begin our analysis by a simple comparison of the Bernoulli, beta-Bernoulli, and Dirichlet-categorical baselines themselves.  For the latter two cases, we use the offset approximations of Equations~\ref{e_ergm_den_approx} and \ref{eq_dcat_apx}, respectively; all parameters were estimated by maximum likelihood, using direct minimization of the model deviance under parameter constraints \citep{byrd.et.al:siamjsc:1995}.  The results are shown in the first three models of Table~\ref{t_ucds}.  The first, Bernoulli, model provides a baseline in which density is taken to be constant across all networks.  The estimated parameter corresponds to an expected density of 0.42, which is (necessarily) equal to the marginal probability of a randomly selected edge being present \emph{when all networks are pooled.}  Edges are not equally likely to be present across networks, however: the mean of the densities of each of the 61 UCDS power networks is 0.34, with the densities themselves ranging from 0.0015 to 0.625 (SD 0.14).  This is not primarily a matter of network size (the correlation of size and density is only about -0.31), suggesting additional sources of heterogeneity.  The second model of Table~\ref{t_ucds} shows a beta-Bernoulli fit to the same data.  The deviance and AIC statistics indicate a substantial improvement versus the Bernoulli baseline, and the  mean (0.33) and standard deviation (0.16) of the expected density are in good agreement with the data despite the use of the offset approximation.\footnote{Note that we do not here employ the constant mean degree version of this model, since inspection of the data reveals that it is the density rather than the mean degree that is closer to constant as size varies.}  




\begin{sidewaystable}
\begin{center}
\begin{tabular}{lrlcrlcrlcrl}
\hline\hline 
                   & \multicolumn{2}{c}{Bernoulli} & & \multicolumn{2}{c}{Beta-Bernoulli} & & \multicolumn{2}{c}{Dirichlet-Categorical} & & \multicolumn{2}{c}{D-C Extended} \\ \hline
Term               & $\hat{\theta}$ (s.e.) & & & $\hat{\theta}$ (s.e.) & & & $\hat{\theta}$ (s.e.) & & & $\hat{\theta}$ (s.e.) &\\ \hline
Edges              & -0.874 (0.027)&$^{***}$ & &               &      & &                 &      & &  & \\   
Mixture ($\alpha$) &               &      & & 2.593 (0.512) &$^{***}$ & & 0.423 (0.076) &$^{***}$ & & 1.176 (5.038$^a$) &$^b$ \\   
Mixture ($\beta$)  &               &      & & 5.377 (1.120) &$^{***}$ & & 2.637 (0.469) &$^{***}$ & &  3.575 (0.205$^a$) &$^{***}$$^b$\\
Mixture ($\gamma$) &               &      & &               &      & & 1.926 (0.350) &$^{***}$ & & 1.843 (0.436$^a$) &$^b$\\
Gender (F$\to$M)   &               &      & &               &      & &               &      & & -0.435 (0.070) &$^{***}$\\
Gender (M$\to$F)   &               &      & &               &      & &               &      & &  0.048 (0.051) &\\
Gender (F$\to$F)   &               &      & &               &      & &               &      & & -0.198 (0.068) &$^{**}$\\
Age Difference     &               &      & &               &      & &               &      & &  0.020 (0.004) &$^{***}$\\
Tenure Difference  &               &      & &               &      & &               &      & &  0.086 (0.026) &$^{***}$\\
Founder Difference &               &      & &               &      & &               &      & &  0.234 (0.053) &$^{***}$\\
Cyclic Triples     &               &      & &               &      & &               &      & & -0.165 (0.015) &$^{***}$\\
Transitive Triples &               &      & &               &      & &               &      & &  0.049 (0.004) &$^{***}$\\ \hline
Null Deviance          & \multicolumn{2}{r}{8875 (6402 df)} & & \multicolumn{2}{r}{8875 (6402 df)} & & \multicolumn{2}{r}{8875 (6402 df)} & & \multicolumn{2}{r}{8875 (6402 df)} \\
Residual Deviance      & \multicolumn{2}{r}{8173 (6401 df)} & & \multicolumn{2}{r}{7013 (6400 df)} & & \multicolumn{2}{r}{6797 (6399 df)} & & \multicolumn{2}{r}{5038 (6391 df)} \\
AIC                    & \multicolumn{2}{r}{8175}  & & \multicolumn{2}{r}{7017}                & & \multicolumn{2}{r}{6802}                & & \multicolumn{2}{r}{5060}\\ \hline
\multicolumn{12}{c}{\footnotesize $^{*}$ $p<$0.05, $^{***}$ $p<$0.01, $^{***}$ $p<$0.001; $^a$ standard error transformed from raw estimate; $^b$ $p$-values based on untransformed estimates} \\ \hline \hline
\end{tabular}
\caption{\label{t_ucds} Fitted models for the Urban Communes data set.}
\end{center}
\end{sidewaystable}

Since we have strong reason to expect asymmetry in this network, we next examine the Dirichlet-categorical baseline.  As expected, this fits substantially better than either of the other two baseline families, and it closely reproduces the rates of mutual, asymmetric, and null dyads (predicted: 0.08, 0.53, and 0.39 vs. observed: 0.06, 0.56, 0.38) even in its approximate form.  Given its markedly better fit to the data (and motivation in terms of variable reciprocity rates), we employ this baseline in subsequent modeling.

Having chosen a baseline family, we can now elaborate it to incorporate other types of structural influences.  For illustrative purposes, we here consider three classes of covariate effects (motivated by prior literature).  First, we consider the hypothesis that perception of personal power in relationships varies by gender, with an overall tendency for males to be more likely to be seen as exercising power over women than vice versa.  We capture this via mixing terms for gender.  We also hypothesize that age, tenure within the commune, and the role of having been a founder of the commune are all bases of power, and hence those with greater age or tenure (or who are founders) will tend to be more likely to been seen as exercising power over those who are younger or newer to the group.  We capture these effects by means of signed difference terms for each pair of individuals (with ``founder'' status coded dichotomously).  Finally, we hypothesize that, net of these factors (and baseline reciprocity), perceived power relations will be subject to both transitive closure and avoidance of cyclic closure (reflecting a locally hierarchical organization of ties).  Thee terms are captured via counts of cyclic and transitive triples, respectively.

To fit the joint model, we employ a custom extension of the \texttt{ergm} \citep{hunter.et.al:jss:2008} package for the \textsf{R} statistical computing system \citep{rteam:sw:2014} that supports pooled model estimation and the baseline mixture terms.  In order to respect the parameter value constraints for the mixture terms, we reparameterize to estimate $\log\left(\alpha-1\right)$ (and likewise for $\beta$ and $\gamma$); these (and their associated standard errors) are back-transformed in Table~\ref{t_ucds} to simplify interpretation.   Models were otherwise fit using standard methods (i.e., MCMC-MLE with MPLE initialization).

The results of this combined analysis are shown in the fourth model of Table~\ref{t_ucds}.  Inclusion of additional effects has not accounted for local asymmetry in the data, as evidenced by the relatively large $\hat{\beta}$ value; holding out other effects, the post-conditioning baseline ``acts like'' a model with mutual, asymmetric, and null probabilities of 0.18, 0.54, and 0.38 (respectively).  We note that the sum of all parameters has increased slightly (from 4.99 to 6.59), suggesting that the inclusion of other effects has accounted for some (but by no means all) of the variation in density and reciprocity across networks.  Turning to our covariate effects, we note that, as hypothesized, males are substantially more likely to send ties to females than vice versa.  Interestingly, there is also a global gender effect, with female-female ties somewhat less likely than male-male ones (the reference category).  Gender mixing in the perceived power network thus has two components: a general tendency of males to be perceived as exercising power relative to females, and a cross-gender power gradient in which males are more likely to be seen as exercising power over females than vice versa.  In addition to gender, we also find positive and significant effects for age differences, tenure in the commune, and being a commune founder (all in the expected direction).  Those who are older, who have been in the group longer, and who are implicated in the group's formation would seem to have an advantage over others.  Finally, we find that, net of all other effects, there remains a tendency for perceived power relations to show transitive closure (and avoid the formation of local cycles).  Interestingly, the former effect is over three times as large as the latter, which suggests that an avoidance of claiming power over those ``above'' is more important than a tendency to seek power over those ``below'' in these networks.

\subsection{Graph Mixtures as ``Contagious'' Process Models} \label{sec_contagion}

As noted in Section~\ref{sec_mix_beta}, some network mixture families arise from easily described behavioral mechanisms.  For instance, consider a ``contagious'' tie formation process \citep[in the behavioral sense of][]{coleman:bk:1964} operating as follows.  Assume a homogeneous set of actors, $V$, each of whom controls his or her outgoing ties.  Time advances in rounds\footnote{This process can easily be embedded in continuous time without essential difficulties; as this provides no particularly useful insight, however, we restrict ourselves to the discrete time case.}, such that at each round a randomly chosen actor re-evaluates his or her relationship with a randomly selected alter.  We assume that decisions to form or dissolve ties depend on idiosyncratic factors and can be treated as random, with the exception that actors' propensities to form or maintain ties are influenced by the current density of the network.                 In particular, consider the case in which ego $i$ is evaluating his or her relationship with alter $j$ at round $t$.  If $y^{(t)}$ denotes the state of the network at time $t$, there are $e((y^{(t)})^c_{ij})$ out of $e^*-1$ possible edges in the graph; we assume that $i$ updates his or her relationship with $j$ as follows:
\begin{equation}
y^{(t+1)} = \begin{cases}1 & \mathrm{if}\quad U_t < \frac{e((y^{(t)})^c_{ij})+\alpha}{e^*-1+\alpha+\beta}\\ 0 & \mathrm{otherwise} \end{cases} \label{e_contagion}
\end{equation}
where $U_t$ is an independent uniform deviate on the $(0,1)$ interval.  Thus, when $i$ observes that ties are extremely common, he or she becomes more likely to create or sustain them.  Likewise, when $i$ notes that ties are rare, he or she reduces his or her own relationship count accordingly.  

The above can be thought of as a very simple process of social influence, in which actors' propensities to engage in social relationships are based in part on the observed behaviors of others.  Such a sideways-looking component to relationship dynamics is plausible in settings for which the presence of a relationship itself is a visible and socially meaningful statement (e.g., friendship in small groups, inter-organizational philanthropic ties, etc.).  Here, we have isolated only this specific mechanism, leaving all other aspects of network dynamics idiosyncratic; this allows us to gain insight into the particular consequences of influence in tie formation per se, holding out other effects.

To examine the behavior of the influence process, we begin by observing that the updating rule of Equation~\ref{e_contagion} corresponds to a Gibbs sampler whose equilibrium distribution (per the results of Section~\ref{sec_mix_beta}) is a beta-Bernoulli graph with parameters $\alpha$ and $\beta$.  The long-run behavior of the process is thus characterized by the exponential family form of Equation~\ref{e_pmf_den_erg}, and equivalently the beta mixture form of Equation~\ref{e_pmf_den}.  From the latter, we can observe that realizations of the network at random times will look like homogeneous Bernoulli graphs with densities drawn from a Beta$(\alpha,\beta)$ distribution; the expected density will thus be equal to $\alpha/(\alpha+\beta)$, with decreasing variation as $\alpha+\beta \to \infty$.  When $\alpha,\beta<1$, the result will be a bimodal distribution that often produces very sparse or very dense graphs, with the system shifting between modes at random times.  Such behavior provides an example of \emph{unstable norm formation}, in which actors are strongly influenced to conform to the current norm (here, the tie rate) but the norm itself can fluctuate over time.

\begin{sidewaysfigure}
\begin{center}
\includegraphics[width=\textheight]{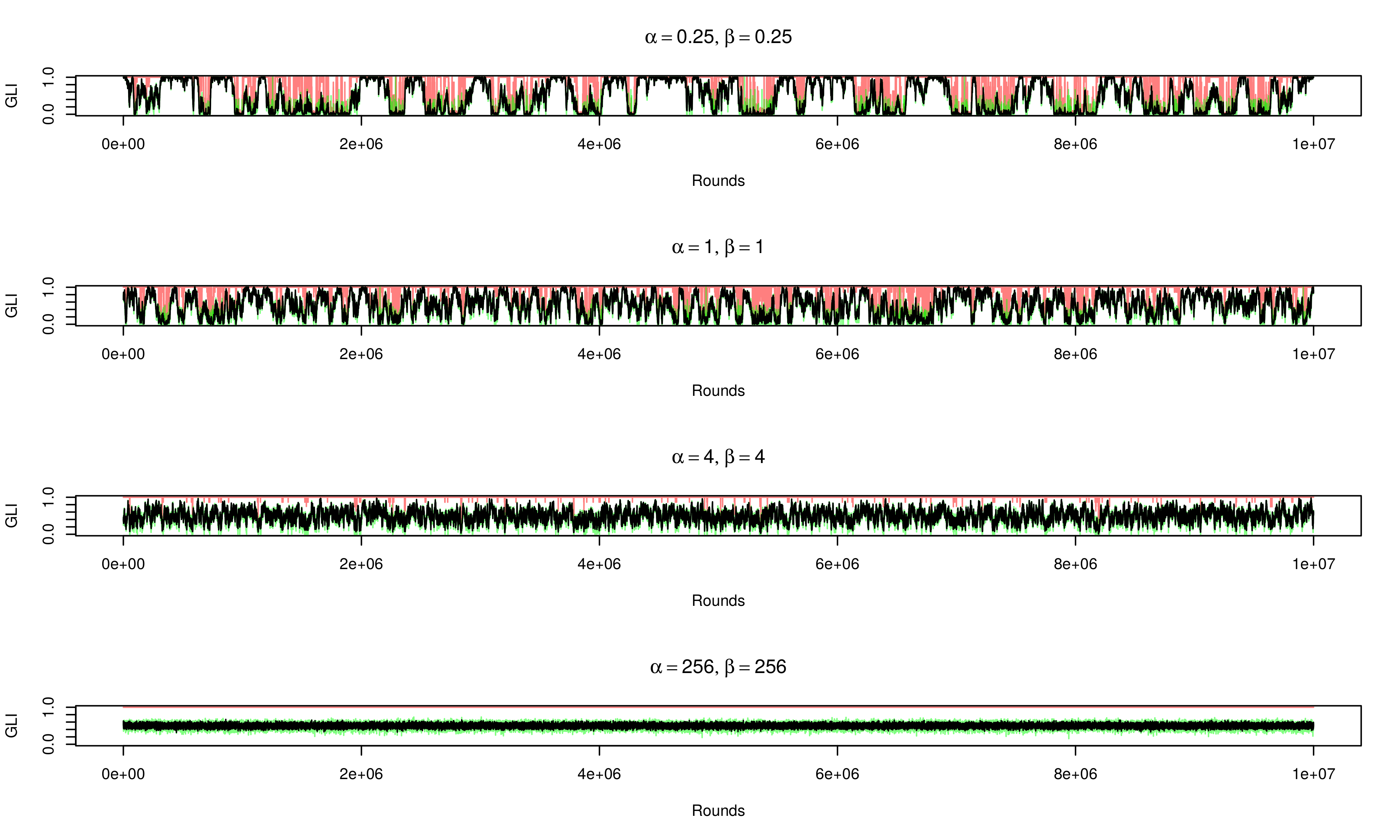}
\caption{\label{f_bbtrace} Illustrative realizations of the contagious tie formation process with $|V|=15$, for various values of $\alpha$ and $\beta$.  Black lines show density, red lines Krackhardt connectedness, and green lines edgewise reciprocity; all runs seeded with an exact draw from the appropriate equilibrium distribution.}
\end{center}
\end{sidewaysfigure}

These fluctuations can be quite dramatic, as illustrated by the top panel of Figure~\ref{f_bbtrace}.  The figure shows summary graph-level indices (GLIs) for a typical realization of the tie formation process on a small group ($|V|=15$) in equilibrium; density is indicated in black, Krackhardt connectedness \citep{krackhardt:ch:1994} in red, and edgewise reciprocity in green (all measures are on a (0,1) scale).  The top panel shows a case ($\alpha=\beta=0.25$) in which the marginal density distribution is strongly bimodal, with the result that the social system tends to spend long periods within either very sparse or dense regimes.  Fluctuations between these regimes can be extremely rapid relative to the lengths of stable periods, and stable periods themselves are punctuated by occasional excursions of very short duration.  Such behavior is a result of the imitative nature of the social process: a random spike (or drop) in social density tends to lead to a cascade of further tie additions (or removals), amplifying the initial random event.  By contrast, as $\alpha$ and $\beta$ are increased (lower panels) such fluctuations begin to be damped out relative to the baseline tie formation propensity, and the system stabilizes.  For $\alpha,\beta \gg 1$, the marginal distribution of density is approximately Gaussian, and the system gradually approaches a homogeneous Bernoulli graph.  By means of the beta mixture representation, we can thus easily characterize the impact of varying levels of influence on the evolution of the social network.

\begin{figure}
\begin{center}
\includegraphics[width=\textwidth]{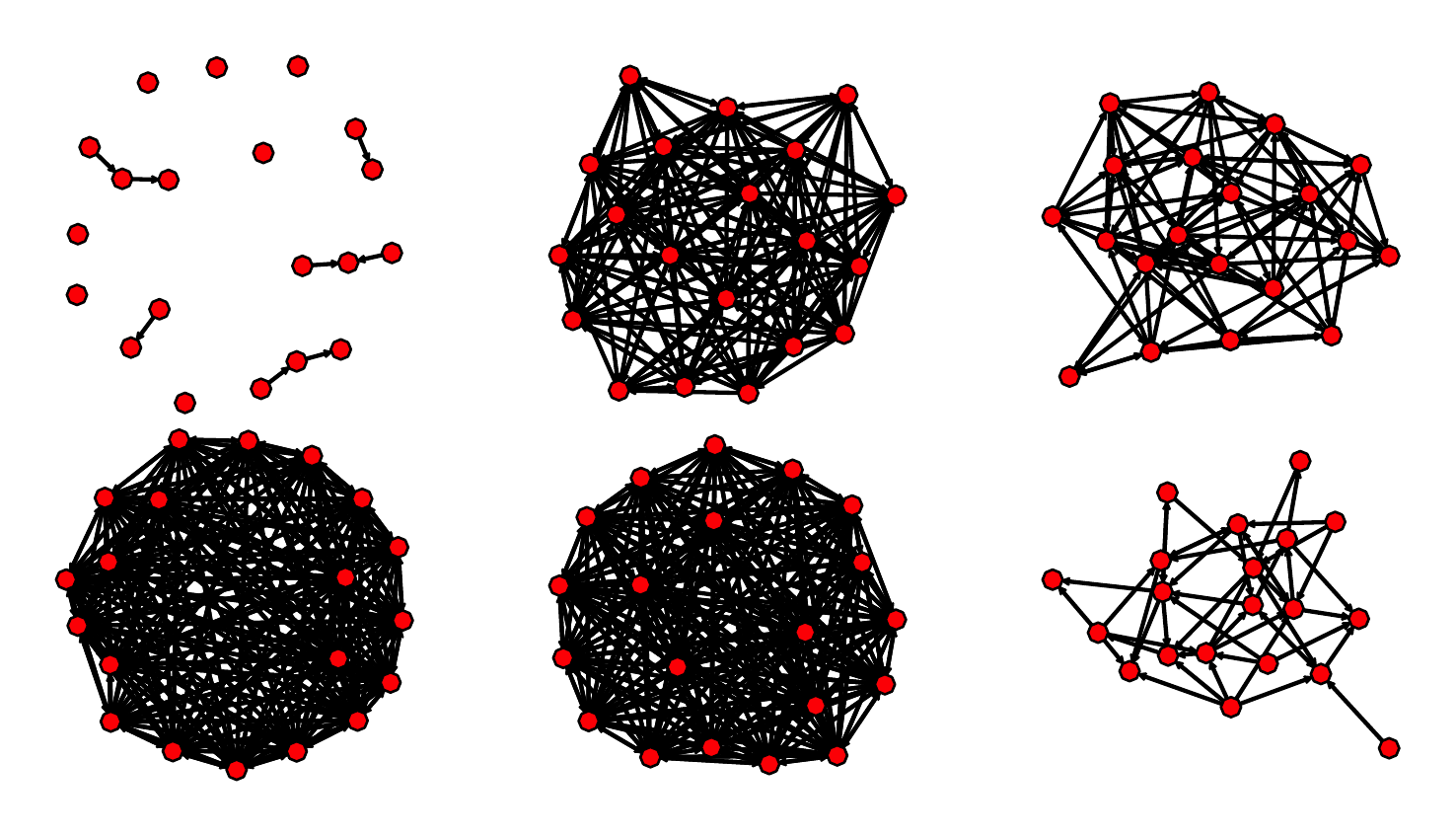}
\caption{\label{f_bbexample} Six independent realizations from the equilibrium distribution of the ``contagious'' tie formation process with parameters $\alpha=\beta=0.5$ ($|V|=20$).  Density and other structural properties vary radically, despite being generated by a common underlying mechanism.}
\end{center}
\end{figure}

The fact that simple influence processes can lead directly to cross-sectional behavior with graph mixture representations provides an important theoretical motivation for studying the latter, and provides an important counterpoint to the conventional argument that differences in observed network properties (either across groups or over time) necessarily reflect the presence of heterogeneity in either process or covariates.  While heterogeneity can certainly lead to differences in network structure, Figures~\ref{f_bbtrace} and \ref{f_bbexample} graphically illustrate that networks with wildly different characteristics can emerge from even a very simple, homogeneous social process.  Indeed, this observation leads to the rather disturbing conclusion that, if such processes are at work in real social networks, the current generation of network research may be drawing very misleading conclusions: most network studies are based on only a single observation, which as we have seen is insufficient to identify graph mixture models of the type arising from the network influence process.  While dynamic network data is becoming more common, the fact remains that most network data collected to date can neither identify nor rule out the presence of mixture-like behavior, \emph{even in cross-section.}  Considerable changes in data collection practice will be needed to appropriate address this phenomenon.

\subsection{Mixture Models as Priors for Network Inference}

The problem of inferring a true network from error-prone data---\emph{network inference}---is of obvious importance for social network researchers.  At least since the pioneering work of Bernard, Killworth, Sailer, and colleagues \citep[e.g.][]{killworth.bernard:ho:1976,bernard.killworth:hcr:1977, killworth.bernard:sn:1979,bernard.et.al:sn:1979, bernard.et.al:ara:1984}, it has been known that most social network data has a non-trivial level of error.  \citet{krackhardt:sn:1987a} introduced a general approach to data collection that facilitates network inference by pooling answers across subjects; statistical methods for network inference in such situations have been developed e.g. by \citet{kumbasar.et.al:ajs:1994} and \citet{butts:sn:2003}.  The latter introduced a fully Bayesian treatment of the network inference problem, with arbitrary ERGM network priors.  As a practical, minimally informative default prior Butts suggests the use of a homogeneous Bernoulli graph distribution, with expected density chosen by prior theory.  As we have seen, the Bernoulli graph concentrates substantial probability mass on graphs with densities near the expectation, and poor choice of prior expected density can thus lead to inferred graphs that are either too dense or too sparse.  While this problem is readily overcome by the large numbers of observations (e.g., 20-35 per dyad) obtained in studies like those of \citet{krackhardt:sn:1987a,krackhardt:asq:1990}, such data collection schemes place a large burden on subjects and are difficult to scale.  This highlights the need for default network priors that work efficiently under a range of densities, while still being minimally informative vis a vis structural details.  The mixture models introduced here are excellent candidates for this application, as they maintain conditional Bernoulli or categorical graph structure while also allowing density to be highly variable.  (In this sense they are analogous to continous mixtures of Gaussian distributions in a more conventional statistical context, which are commonly used as robust alternatives to simple Gaussians \citep{gelman.et.al:bk:2003}.)

To evaluate the potential of baseline mixtures as potential priors for network inference models, we here conduct a simple computational experiment in which we construct true (criterion) graphs with varying levels of density and reciprocity and attempt to infer them from error-prone data using the methods of \citet{butts:sn:2003}.  We draw criterion graphs of moderate size ($|V|=50$) from $U|man$ distributions with respective expected density levels of 0.05, 0.25, and 0.5, and with expected edgewise reciprocity levels of 0.05, 0.5, and 0.95 (a three-by-three design, with nine conditions).  For each density/reciprocity condition, we draw 250 independent criterion graphs to serve as targets for network inference.

Given each criterion graph, we generate a series of error-prone ``observations'' in the general style of cognitive social structure data \citep[aka CSS data, see][]{krackhardt:sn:1987a}.  As a simple model of reporting error, we assume that each data source generates false positive errors at random with probability 0.05 and false negative errors with probability 0.5; i.e., given the presence of an $(i,j)$ edge in the criterion graph the source will fail to report it 50\% of the time, and given the absence of an $(i,j)$ edge in the criterion graph the source will falsely report one on 5\% of occasions.  (These error rates are chosen to reflect the much higher rates of false negative versus false positive errors in most human informant data, and roughly follow those observed by \citet{butts:sn:2003}.)  15 sets of observations (CSS ``slices'') are then generated for each criterion graph, using the above error rates.

Given our simulated observations, we finally seek to infer the criterion graph using the Bayesian model of \citet[section 2.5; i.e., multiple observers and general ERG priors]{butts:sn:2003}.\footnote{In all cases, independent Beta(1,11) priors were used for informant error parameters; posterior inference was based on the edgewise marginal mode of 300 draws taken from 3 independent MCMC chains following 100 burn-in iterations (convergence verified using the Gelman-Rubin scale reduction diagnostic).  All analyses performed using \textsf{sna} \citep{butts:jss:2008b} and \textsf{statnet} \citep{handcock.et.al:jss:2008}.}  As our interest is in evaluating the effectiveness of baseline mixtures as network priors, we test four different prior distributions: a uniform Bernoulli graph with relatively low expected density (expected density 0.05); a uniform Bernoulli graph with relatively high expected density (expected density 0.5); a beta-Binomial graph with mixing distribution Beta(0.5,0.5); and a Dirichlet-categorical graph with mixing distribution Dirichlet(0.5,0.5,0.5).  While all of the above can be considered to be weakly informative priors, the Bernoulli graphs in practice put substantial weight on their expected densities, and are hence inefficient when the criterion graph density is far from the prior expectation.  To the extent that mixture priors rectify this problem, they should produce higher levels of posterior accuracy from smaller numbers of observations.

\begin{figure}
\begin{center}
\includegraphics[width=\textwidth]{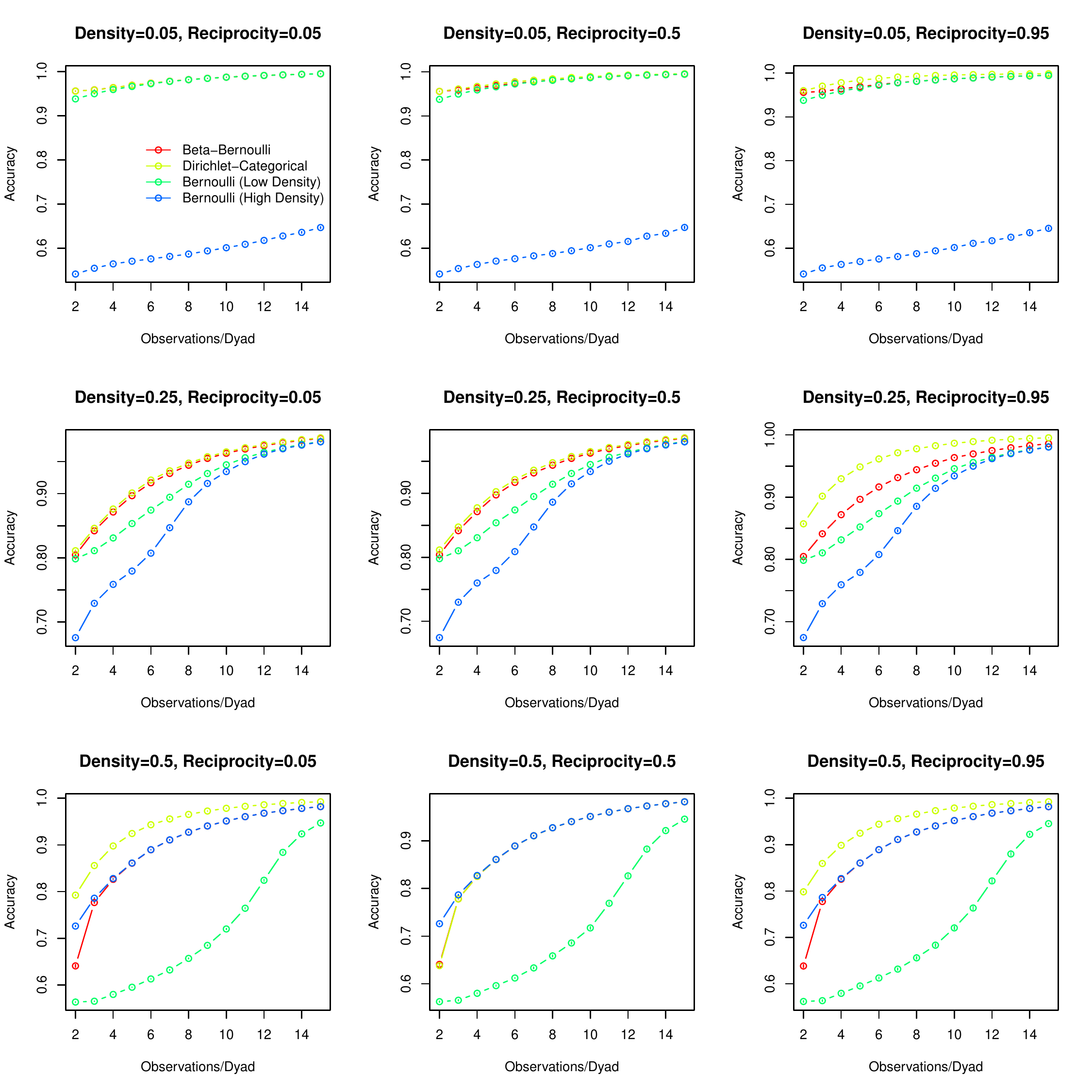}
\caption{\label{f_hamacc} Hamming accuracy (fraction of edge states correctly classified) by number of observations per dyad, for criterion graphs with varying expected density and reciprocity values.}
\end{center}
\end{figure}

The overall impact of network prior on inferential accuracy is summarized in Figure~\ref{f_hamacc}.  Each panel of Figure~\ref{f_hamacc} shows the average Hamming accuracy (fraction of directed dyads correctly classified) of each network inference model for the 250 criterion graphs in the respective condition.  To assess the impact of network prior on efficiency, we repeatedly fit each network inference model to data subsets of increasing size, varying the number of observations per dyad (CSS slices) from 2 to 15.  Unsurprisingly, extremely high levels of accuracy are possible with large numbers of observations; for the Bernoulli graph priors, however, criterion graphs with densities far from the prior density are poorly estimated with even moderate numbers of observations per dyad.  By contrast, the graph mixture models are consistently strong performers, working well across a range of densities and with small numbers of observations.  In particular, the Dirichlet-categorical prior proves especially effective, often allowing greater than 90\% accuracy with as few as 3--5 observations per dyad.  This superiority is most noticeable at extreme reciprocity values, where the Dirichlet-categorical prior generally outperforms even a well-calibrated Bernoulli prior.  The advantage of the Dirichlet-categorical model in these cases is due to its ability to pool information within dyads: when, e.g., reciprocity is inferred to be extremely high, the presence of a confidently inferred $(i,j)$ edge allows the model to infer that a $(j,i)$ edge is also likely (similar arguments holding, mutatis mutandis, in the low-reciprocity case).  While strong reciprocity biases improve performance with this prior, the lack of such biases (e.g., bottom center panel) does not substantially degrade it.  The Dirichlet-categorical model thus appears to be both efficient and robust, in comparison with simple Bernoulli or even Bernoulli mixture priors.

\begin{figure}
\begin{center}
\includegraphics[width=\textwidth]{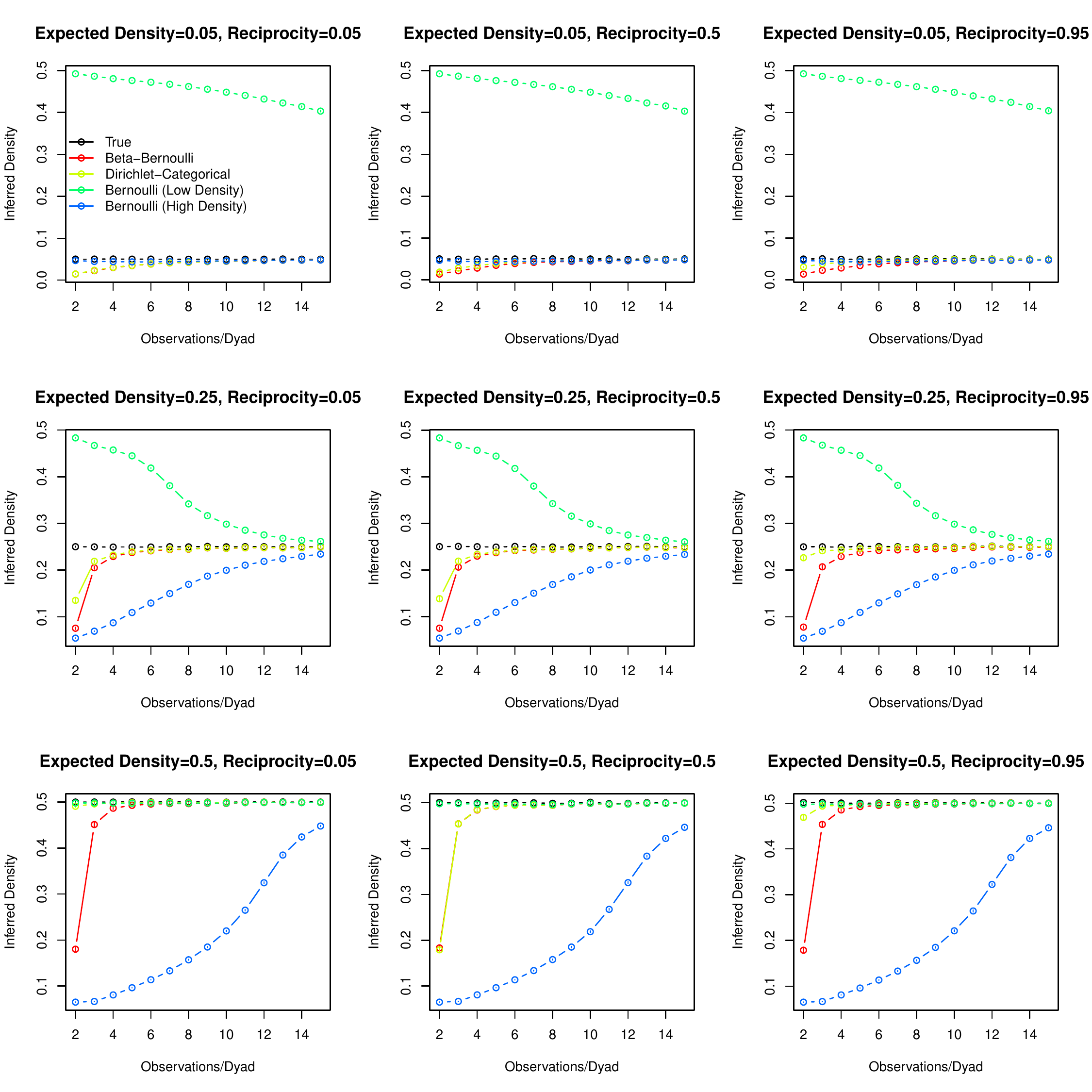}
\caption{\label{f_infden} Inferred density by number of observations per dyad, for criterion graphs with varying expected density and reciprocity values.}
\end{center}
\end{figure}

Another view of the benefits of network mixture priors is provided by Figure~\ref{f_infden}, which shows inferred versus expected criterion density for each experimental condition.  Recalling that the principal problem with the simple Bernoulli graph prior is that it places excessive prior mass on graphs with a particular density, we see immediately that many observations (often more than the 15 used here) are required to correctly infer the density of the criterion graph under Bernoulli graph priors whose expected densities are far from the true value.  By contrast, both mixture prior families produce inferences that quickly converge to the true density, regardless of value.  Indeed, we see excellent convergence with 4--5 observations under both mixture priors, with the Dirichlet-categorical showing excellent performance with as few as 2--3 observations per dyad in some conditions.  Since network density drives a wide range of other structural characteristics \citep[see, e.g.][]{anderson.et.al:sn:1999,butts:sn:2006,faust:sm:2007}, accurately inferring it is critical for many practical applications.

To summarize, examination of the performance of network mixture models as weakly informative priors for network inference suggests that they can provide substantial robustness and efficiency gains versus Bernoulli graph priors.  Although the latter perform well when the expected prior density is close to that of the criterion graph, they can be extremely inefficient in other cases---a serious concern in typical settings, for which there may be minimal ex ante guidance regarding graph density.  By contrast, both beta-Bernoulli and Dirichlet-categorical graphs with diffuse mixture components work well across a range of criterion densities, converging with even a small number of observations per dyad.  Moreover, models such as the Dirichlet-categorical can outperform Bernoulli priors by pooling information within dyads, exploiting reciprocity bias (common to most social networks) to use more confidently observed edges to help predict the states of more ambiguous ones.  Since there seems to be little penalty associated with their use, we suggest network mixture priors as a superior alternative to standard Bernoulli priors as a ``default'' choice for typical network inference settings.

\section{Conclusion}

In this paper we have explored the use of continuous mixtures of baseline graph distributions---specifically, Bernoulli and $U|man$ graphs---as graph distributions for social network applications.  We examined in detail the special cases of the beta-Bernoulli and Dirichlet-categorical graphs, distributions with dyadic dependence that nevertheless have analytically and computationally tractable forms.  Inference for these families is straightforward, but as noted depends upon having multiple observations; this is an immediate consequence of the fact that these families capture cross-graph outcome heterogeneity, which cannot be assessed from a single observation.  The baseline mixture models are potentially useful as null models alongside their Bernoulli and $U|man$ counterparts, as starting points for elaboration (in exponential family form), or for a variety of other purposes.

One interesting and possibly arresting consequence of the developments of this paper is the observation that graph distributions with non-trivial behavior can arise from even simple dynamic processes, and that even the cross-sectional behavior of such processes cannot be characterized from a single realization.  Given that the vast majority of social network data is in the form of single, cross-sectional observations, it follows that most extant data is insufficient to allow for the detection of such processes (much less to infer their associated parameters).  Data collection designs obtaining multiple observations of the same system over time, or alternately observations of multiple systems with comparable underlying behavior, would greatly alleviate this problem.

On the bright side, our results also show that using baseline mixture models as network priors can greatly \emph{decrease} data requirements for network inference problems.  Indeed, we find that as few as 5 or so observations per dyad may be enough to obtain good performance under realistic error rates; this is far smaller than the 20--35 observations per dyad typical of full CSS designs, and much easier to collect.  The mixture priors studied here appear to improve both efficiency and robustness relative to Bernoulli priors under a range of conditions, and we recommend them as superior for typical interpersonal network settings.

Given the versatility and importance of continuous graph mixtures, further theoretical development in this area seems warranted.  In particular, it would be useful to be able to characterize which types of network dynamics lead to cross-sectional mixture behavior of the form considered here, and hence what sorts of heterogeneity may be expected across networks in particular empirical settings.  Without such an understanding, it becomes easy to ascribe to covariate effects what may in fact be due to natural variability.  Mixture models are also a natural basis for building models for networks with unobserved heterogeneity, of course, and further theoretical developments will be helpful here as well.  It is hoped that this paper provides an initial basis for pursuing these and other related questions.

\bibliography{ctb}


\end{document}